# Anomaly detection models for IoT time series data

Federico Giannoni
Mtr: 187406
federico.giannoni@studenti.unitn.it

Marco Mancini
Mtr: 187403
marco.mancini@studenti.unitn.it

Federico Marinelli
Mtr: 187419
federico.marinelli@studenti.unitn.it

*Data Mining - Project Report*

## 1. INTRODUCTION

In-situ sensors and Wireless Sensor Networks (WSNs) have become more and more popular in the last decade, due to their potential to be used in various applications of many different fields. As of today, WSNs are pretty much used by any **monitoring** system: from those that are health care related, to those that are used for environmental forecasting or surveillance purposes.

All applications that make use of in-situ sensors, strongly rely on their correct operation, which however, is quite difficult to guarantee. These sensors in fact, are typically cheap and prone to malfunction. Additionally, for many tasks (e.g. environmental forecasting), sensors are also deployed under potentially harsh weather condition, making their breakage even more likely. The high probability of erroneous readings or data corruption during transmission, brings up the **problem of ensuring quality of the data collected by sensors**. Since WSNs have to operate continuously and therefore generate very large volumes of data every day, the quality-control process has to be automated, scalable and fast enough to be applicable to streaming data.

The most common approach to ensure the quality of sensors' data, consists in automated detection of erroneous readings or anomalous behaviours of sensors. In the literature, this strategy is known as **anomaly detection** and can be pursued in many different ways.

## 2. RELATED WORK

Algorithms for anomaly detection in sensor's time series data, can be subdivided in the following macro-classes:

- **Statistical methods**; these methods use past measurements to approximate a model of the correct behaviour of a sensor (or of whatever component we are trying to monitor). Whenever a new measurement is registered, it is compared to the model and, if it results to be statistically incompatible with it, then it is marked as an anomaly. Statistical methods can not only be applied on single readings [1], but also on windows of readings [2]. A window-based approach typically helps in reducing the number of false positives. An example of a very common statistical anomaly detection method, is the so-called **low-high pass filter**, which classifies readings as anomalies based on how different they are from the running average of past measurements.
- **Probabilistic methods**; these methods revolve around the definition of a probabilistic model that can either be parametric or non-parametric (depending on whether or not the sensors' measurements follow a well-known distribution or not). Classification of anomalies is then performed by measuring the probability of a reading with respect to the model. If the probability falls below a predefined threshold, then it is labelled as an anomalous event. The models defined by probabilistic methods can be very simple, but also very complicated, possibly encoding relations between measurements through time, using either Hidden Markov Models (HMMs) [3] or Bayesian Networks (BNs) [4]. Methods based on BNs and HMMs however, are typically very computationally expensive and do not scale well, especially with streaming data.
- **Proximity-based methods**; these methods rely on distances between data measurements to distinguish between anomalous and correct readings. A very famous proximity-based algorithm is the Local Outlier Factor (LOF) [5], which assigns an outlier score to each reading ri based on the density of measurements around its k nearest neighbours and the density of measurements around ri. Readings with high outlier scores are labelled as anomalies.
- **Clustering-based methods**; these methods are a subset of proximity-based algorithms. Here, measurements are first used to create clusters. Then, new measurement that are assigned to small and isolated clusters or measurements that are very far from their cluster's centroid, are labelled as anomalous [6].
- **Prediction-based methods;** these methods use past measurements to train a model that can predict the value of the next measurement in the sensors' time series data. If the actual measurement is too different from the predicted one, then it is labelled as anomalous. There are many prediction-based algorithms for anomaly detection, some are based on very simple machine learning models, such as 1-class SVMs [7], while others are way more complex and make use of Deep Neural Networks (DNN) with Long Short-Term Memory (LSTM) cells as their predictive model [8].

Obviously, deciding which algorithm to use strongly depends on the type of data to be monitored. For instance, clustering approaches and proximity based approaches do not work well with high dimensional data, as distances between points tend to dilate as the number of dimensions increases. Similarly, techniques based on HMMs and BNs can not efficiently handle multivariate data and are therefore suitable for 1-dimensional samples only.

## 3. A CONCRETE CASE STUDY

Our objective was to design and develop an anomaly detection algorithm for sensors deployed in the ETC wastewater plant in Riccione. A synthetic map of the plant is shown in Figure 1. As it is possible to see, the wastewater is pumped from the same source into three separate tanks. In each tank, a trace of four sensors is used to monitor the concentration of oxygen, ammonia, suspended solids and nitrates in the water. Valves and blowers are also present and piloted by a central PLC controller, based on the sensors' readings. The controller can open the valves and manipulate the blowers' power to trigger certain chemical reactions and adjust the level of the compounds that are present in the water. The correct operation of the system strongly depends on the quality of sensors' readings, since the controller manipulates the actuators based on the current advertised state of the water.

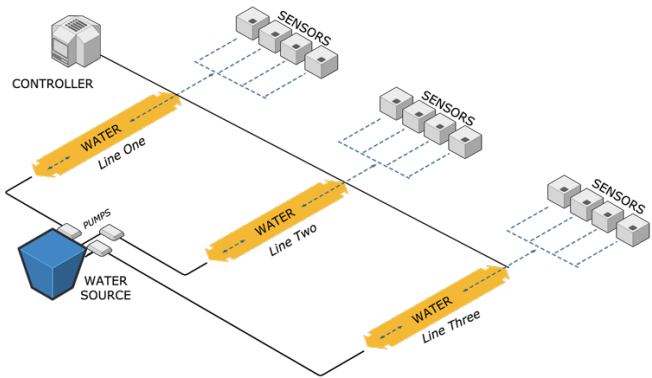

*Figure 1  Schema of the ETC Riccione wastewater plant. Each line has four sensors that register the levels of ammonia, oxygen, nitrates and suspended solids in the respective tanks. All lines are connected by pipes to a controller, which can open and close valves to regulate the level of oxygen in each tank.*

Given their particular importance, we decided to focus on the ammonia sensors. Our objective was to ensure the quality of the data they recorded, by **detecting any potential anomaly and then use the event frequency as a flag to detect** (possibly preemptively) any **fault** (e.g. drifting) or **breakage** of the sensors. Formally, given a time series of readings of an ammonia sensor, $x_1, x_2, \ldots, x_i, \ldots, x_N$, (where the subscript represents the instant in which the reading was recorded), we are interested in detecting all anomalous events $X$. Additionally, fixed a time interval $\Delta t$, we would like to determine the number of anomalies $m$, such that, if $\{x_i, \ldots, x_i + \Delta t\} \in X$ and $|\{x_i, \ldots, x_i + \Delta t\}| \geq m$, then $P(\text{sensor breakage}) \geq M$, with $M$ as close as possible to 1. The following Sections report our thought process and our plan of action to solve the aforementioned problem.

## 4. DATA ANALYSIS AND PRE-PROCESSING

Choosing which algorithm to use obviously requires a certain level of knowledge of both the problem and the data. In order to understand what could be used to identify anomalies in the ammonia sensors, we first had to analyze the dataset that was provided to us by the people at ETC Riccione.

The original dataset consisted of several CSV files, each containing one hour of acquisitions. However, we reorganized it to obtain a separate CSV file per sensor (e.g. one file for all acquisitions of the ammonia sensor on trace one, another for all acquisitions of the ammonia sensor on trace two and so on).. Each line of these new CSV files contained a sensor reading, together with its acquisition timestamp, the name of the sensor and an integer value. Acquisitions were sampled with a per-minute frequency, for a total of 1440 acquisitions per day. The dataset covered the period from the beginning of April 2016, to the beginning of July 2017.

### 4.1  Data Cleaning

We began by applying a threshold-based noise filter to remove faulty sensors' readings. The filter worked by identifying and removing all measurements that were either NaN, occurred very rarely (e.g. values of the order of 1032), or did not make sense (e.g. negative values).

Once all sudden spikes, NaNs and negative values were removed, we had to face another issue, this time with the acquisitions' timestamps. Some of the timestamps in fact, were not unique, due to the fact that on the 31st of October, the switch to standard time pushed the clock back one hour from 3:00 AM to 2:00 AM. This lead to duplicate keys in the sensors' data time series. To fix this problem and make the time series usable, we decided to condense the two hours of acquisitions after 2:00 AM of the 31st October, into a single hour, therefore removing the repeating timestamps. This was done by condensing pairs of acquisitions taken in two consecutive minutes $t_i$ and $t_{i+1}$, turning them into a single one represented by their average $\frac{1}{2}(t_i + t_{i+1})$.

### 4.2  Data Visualization

Once our dataset was cleaned, we proceeded with data visualization. We plotted the sensors' time series and their values' distributions and analyzed them. We decomposed the ammonia acquisitions time series and looked at their trend, seasonal and residual (shown in Figure 2). We also ran a Dickey-Fuller test on it and discovered that it presented stationary properties (meaning that there was a repeated behaviour with respect to calendar time). More precisely, we identified **similar patterns in ammonia acquisitions relative to the same month and season**. This informed us that time of acquisition could prove to be a valuable additional feature for our anomaly detection algorithm. Additionally, we also noticed that the residual of the ammonia time series followed a normal distribution.

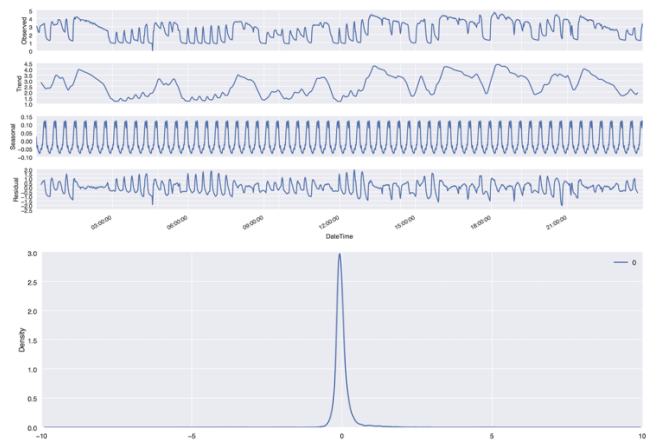

*Figure 2  Above the decomposition of a time series of ammonia acquisitions (sampled by the ammonia sensor of trace three, on day 18-10-2016) . Below we have the distribution of the residual, which is normal. The Dickey-Fuller test results for the time series in the figure were the following:  Test Statistic -5.385561; p-Value 0.000004; # Lags Used 18.0;  Number of Observations Used 1421.0; Critical Value (5%) -2.863576;  Critical Value (1%) -3.434960; Critical Value (10%) -2.567854. Since the test statistic has a negative value that surpasses the critical values, the null hypothesis of a unit root is rejected while the alternate hypothesis of data stationarity is accepted instead.*

Moving on to the distribution of recorded values, we were interested in seeing whether or not these acquisitions approximately fit any well known model, which would have made it possible to apply a parametric probabilistic algorithm for anomaly detection. We therefore plotted histograms for the cleaned acquisitions of oxygen, ammonia and nitrate sensors. Histograms for densities of ammonia readings are shown in Figure 4. As it is possible to see, values registered by ammonia sensors are **approximately normally distributed**, which means that a univariate-based probabilistic approach would make a decent candidate for our detection algorithm.

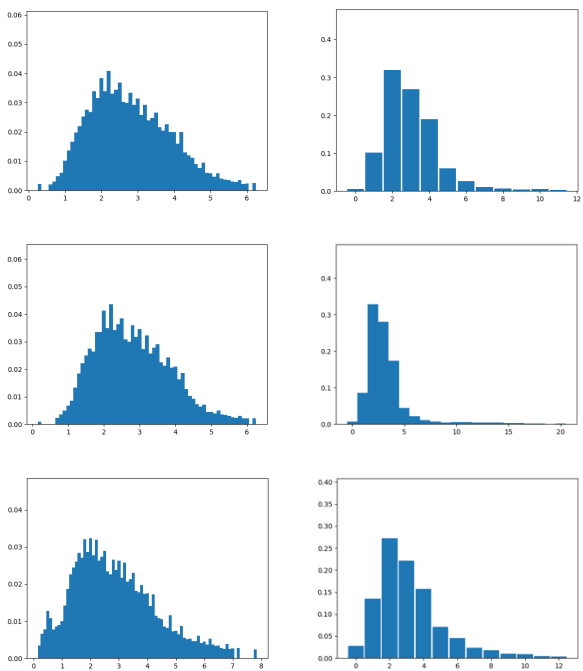

*Figure 3 Histograms for ammonia sensors' readings densities. On the left we have densities for readings rounded to 1 decimal digit, on the left densities for readings rounded to integers (smoothed). Top row is for ammonia sensor in trace one, middle for ammonia sensor in trace two, bottom for ammonia sensor in trace three. As we can see, all histograms have a single peak value and the vast majority of readings are concentrated around said peak.*

## 5. UNCOVERING DATA CORRELATIONS

To understand if there were other measurements we could use as informative features to detect anomalies in the ammonia sensors' acquisitions, we ran some statistical tests on our dataset, in an attempt to discover whether or not there were any correlations between ammonia levels, oxygen level, nitrate levels and suspended solids levels.

We began our analysis with a few simple statistical correlation tests, such as the Pearson correlation test, the Spearman test and the Kendall test. The tests were run on the following pairs of readings: ammonia and oxygen; ammonia and nitrates; ammonia and suspended solids. The correlation factors for ammonia and nitrates was found to be very weak (approximately 0.015 Kendall's value, -1.9e-06 Pearson's correlation coefficient and 0.02 Spearman's value). Acquisitions of ammonia and suspended solids were also found to be statistically independent by our tests (0.03 Kendall's value and 0.04 Spearman's value). The only significant (yet still weak) dependency was found between ammonia and oxygen readings (approximately 0.25 Kendall's value and 0.38 Spearman's value[1]).

---

[1] The Kendall test assigns a correlation value between 0 and 1 to two time series and so does the Spearman test (normally, a Spearman's value below 0.39 is representative of a weak correlation, while a value above 0.59 indicates a strong correlation). The Pearson correlation coefficient is instead a number between -1 and 1, with -1 denoting maximum inverse correlation and 1 maximum direct correlation.

We also ran a causal discovery algorithm called PCMCI [13] on the sensors acquisitions' time series of ammonia, oxygen, nitrates and suspended solids. PCMCI produces a causal time series graph given a dataset sample. This graph (visible in Figure 4), represents the **spatial-temporal dependency structure from which causal pathways can be read off**. In particular, the nodes of a time series graph are defined as the variables at different times and a link exists if two lagged variables are not conditionally independent given the whole past process. Assuming stationarity, the links are repeated in time. The parents P of a variable are defined as the set of all nodes with a link towards it (blue and red boxes in Figure 4). The color of the link represents the correlation value, the more intense it is, the higher is the correlation between the nodes. A link could be either blue or red depending whether its correlational value is positive or not. From the Figure, we can see how the ammonia values are correlated with the oxygen values, which in turn are correlated with the values registered by the nitrates sensors.

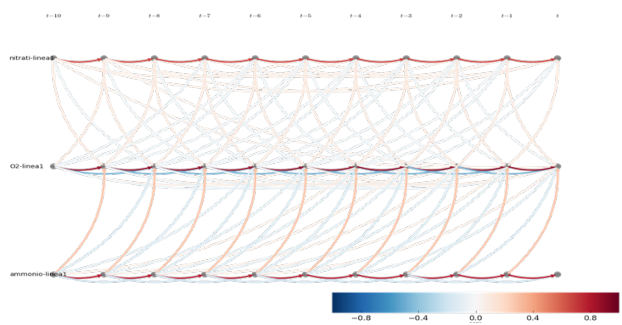

*Figure 4 Spatial-temporal causality graph between ammonia (bottom line), oxygen (mid line) and nitrates (top line) obtained from the PCMCI test. Ammonia and oxygen acquisitions have a greater correlation than oxygen and nitrates, if we look at the intensity of the links' colors. This implies that there is some sort of causal correlation between O2 and NH4 readings.*

To make sure not to miss anything, we also asked some of the experts working at ETC Riccione to explain to us how the levels of each compound influences the others. They confirmed that oxygen and ammonia levels are somewhat correlated (and the same goes for ammonia and nitrates), because oxygen levels are responsible for triggering certain chemical reactions that increase or decrease ammonia concentration in the water. More specifically, when the oxygen is above a certain threshold, the ammonia concentration in the water decreases and vice versa, when the oxygen is low, the ammonia is higher. We concluded that these correlations were not properly captured by the Pearson, the Spearman and the Kendall tests due to the fact that they are trend-based correlation tests, while the correlation between the compounds was threshold-based (meaning that it is not necessarily true that oxygen and ammonia increase/decrease together, but it is true that when the oxygen is above certain levels, the ammonia is low and vice versa).

Our findings informed us that **oxygen and nitrate levels could be used as additional features** for anomaly detection in the ammonia acquisitions. We concluded that these features could not only help in identifying when anomalies are being registered, but also in discriminating false positive events from true positive ones (for instance, a very low ammonia reading may be considered anomalous by itself, but if the oxygen levels at that same time are very high, then it could be that the ammonia concentration truly is that low, given the fact that the chemical reactions triggered by high oxygen levels are taking place).

# 6. EXPLORED SOLUTIONS

Instead of limiting ourselves to develop and fine-tune a single algorithm, we decided to explore different approaches. The following subsections will be dedicated to the description of our solutions. Results will be discussed in the next Section.

## 6.1 Running Average Low-High Pass Filter

The first solution that we implemented was a fairly simple one, yet widely used in the field of sensors' anomaly detection. The basic idea behind it, is to make use of the running average computed based on the last W acquisitions (where W is a sliding window of fixed size), as a means to identify anomalies. More precisely, new acquisitions that differ too much from the running average, are registered as anomalous events. We implemented two different versions of this algorithm, **one for online detection and one for offline detection**.

We start by describing the offline detection version, which is the simpler one. Here, we assume that the entire time series is already available to the algorithm, which can compute its standard deviation. Then, the algorithm computes the running average for each new acquisition $x_i$, based on that acquisition and the previous W - 1. Each $x_i$ is registered as anomalous, if its distance from the running average is greater than the standard deviation previously computed. Pseudo-code for the entire algorithm is shown below.

```
Initialize:
X ← { x1 , … , xn }
W ← window_size
series_std ← std(X)
window ← array[W]

Algorithm:
foreach xi in X do:
   if window is full:
      window.remove(0)
   window.append(xi)
   if (xi · α > average(window) + series_std)  OR
      (xi · α < average(window) - series_std):
      register_anomaly(xi)
```

Where alpha is a hyper-parameter (between 0 and 1) that regulates the algorithm sensitivity to anomalies (the larger alpha, the greater the number of readings that will be listed as anomalies).

The offline version of the algorithm works similarly to a software-based low-high pass filter, cutting off all readings that are either above or below a certain threshold, computed based on the running average and standard deviation in the series.

The online version is fairly similar. The only difference is that we assume that our algorithm does not have access to the whole series a-priori, therefore it cannot compute the standard deviation as previously shown. Instead, an approximated standard deviation is computed at each iteration, based on the acquisitions in the window.

Both versions of the algorithm work exclusively on univariate data, meaning that no additional feature can be used, unless feature transformation is applied.

## 6.2 Univariate Gaussian Predictor

As discussed in Section [Data Visualization], ammonia readings followed a distribution that is more or less approximable to a Gaussian one. We therefore tried to see if a Gaussian probabilistic model could be used to identify anomalous behaviours.

The Univariate Gaussian Predictor [9] is a fairly simple one. It is trained on a time series and uses Maximum Likelihood Estimation (MLE) to approximate the sufficient statistics of the Gaussian model, which are the mean and variance. Once these are computed, the model classifies each new acquisition $x_i$ based on the probability of that value in the distribution $p(x_i)$. A threshold has to be set to decide which probabilities are too low to be considered non-anomalous. Following is the pseudo-code for the algorithm.

```
Initialize:
Xtrain ← { x1, …, xn }
μ ← 1n ∑i = 1n xi
σ2 ← 1n ∑i = 1n (xi-μ)2

Algorithm:
foreach new sample zi do:
   if p(zi) < eps:
      register_anomaly(zi)
```

This algorithm can be run either on windows of acquisitions[2] or on single ones. The main issue of the Univariate Gaussian Predictor is that the **probability of each acquisition is computed as if it was independent from all the previous ones**, which is obviously not the case. To relax the independence assumption one would have to rely on either HMM-based models or BN-based models, which, however, are much more demanding in terms of computational resources.

## 6.3 Seasonal ESD Algorithm

Other than the low-high pass filter, we also tested another statistical-based method, called Seasonal Extreme Studentized Deviate (S-ESD) [10]. This algorithm was released by Twitter in 2017 and can be applied **only on time series that have their residual component symmetrically distributed**. In our case, the constraint is met, since the residual of the ammonia acquisitions series was found to be normally distributed as shown in Figure 2

S-ESD starts by decomposing a time series X. First, its seasonal component S and its trend T are extracted. Then, the median X* is computed from the series' trend. The median can be seen as a stable approximation of the trend. At this point, the residual R is computed from X, S and X* as R = X - S - X*. Using the median instead of the actual trend in the computation of the residual makes it so that the residual does not present any spurious anomalies (such as spikes or negative acquisitions), which are not interesting anyway.

Once the residual is computed, S-ESD uses the Generalized Extreme Studentized Deviate (ESD) statistical test [3][11] to identify any anomalous events based on the value of R and the other series' components and k, the upper bound on the number of anomalous events in the provided series. Below, the pseudo-code of the algorithm is provided.

---

[2] In this case the probability of the window is computed as the product of probabilities of all the acquisitions in the window.
[3] In short, the ESD test matches the null hypothesis that no outlier are presents, against the alternate hypothesis that at most k outliers are present, where k is provided to the algorithm prior to its execution. ESD computes a value $V_j$ = $max_i( |x_i - μ| ) / σ^2$ and a critical value $λ_j$ for k samples, where μ and $σ^2$ are the mean and variance of the series. $V_j$ and $λ_j$ are then used to find the outliers.

```
X ← { x1, …, xn }
S ← extract_seasonal(X)
T ← extract_trend(T)
X* ← median(T, X)
R ← X - S - X*

anomalies ← ESD(R, X, X*, S, T, k)
```

Just like all the algorithms that we described up to now, this one can also only be used on univariate data. Additionally, the upper bound on the number of anomalies (k) has to be fixed a-priori as a hyper parameter.

## 6.4 Local Density Cluster-Based Outlier Factor

The final algorithm that we decided to implement is a cluster-based one that can be used on multivariate data, therefore allowing us to use additional features to discover anomalies in the ammonia acquisitions. The Local Density Cluster-Based Outlier Factor (LDCOF) [12] is an extension of the Cluster-Based Local Outlier Factor (CBLOF) algorithm. CBLOF works by clustering samples using any available clustering algorithm (e.g. k-means) and then divides the clusters into small and large ones. Outlier scores are assigned to new samples xi based on the distance between xi and the closest large cluster centroid. LDCOF is based on the same principle, but it also borrows from the proximity-based Local Outlier Factor (LOF) algorithm, to apply the local density principle when assigning outlier scores, which helps in identifying small clusters as outlying clusters.

To make things clearer, LDCOF first applies a clustering algorithm to a training dataset and it then makes use of two coefficients and (which are hyper-parameters to be tuned), to separate large clusters from small ones. At this point, an average distance from centroid distavg(Ci) is computed for each large cluster Ci. These distances are used to approximate the densities of large clusters. If a sample is placed into a small cluster, its outlier score is computed as the ratio between its distance to the closest large cluster centroid and the average distance of points in that large cluster, from their centroid. On the other hand, if a sample is placed into a large cluster, its outlier score is simply computed as the ratio between its distance from the cluster's centroid and the average distance of points in the cluster from the centroid.

Formally, the outlier score for each new sample p is defined in the following way:

- min( d(p, Ci) )distavg(Ci) if pCj where Cj is a small clusters and Ci are large clusters
- d(p, Ci)distavg(Ci) if pCi where Ci is a large cluster

While the distinction between large and small clusters is performed first by sorting the K clusters so that |C1| |C2| … |CK| and then, given and , we define b as the boundary between large and small clusters, if any of the following conditions hold:

- |C1| + |C2| + … |Cb| ·|D|, where |D| is the size of the training set
- |Cb| / |Cb-1|

Based on the outlier score, a point will be classified as outlier or not. Typically, since outlier scores are ratios in LDCOF, a value above 1.0 would indicate an anomaly, but this may not always be true and can therefore be tuned accordingly to how the data is distributed. Below is a sketch of the pseudo-code of the entire algorithm.

```
X_train ← { x_1, …, x_n }
Model ← Kmeans.fit(X, K)
% cluster training samples

SC, LC ← separate_large_and_small(Model)
% separate large and small clusters as discussed

foreach new sample z_i do:
    cluster ← Model.fit(z_i)
    if cluster in SC then assign_SC_outlier_score(z_i, Model, cluster, SC, LC)
    else assign_LC_outlier_score(z_i, Model, cluster, SC, LC)
```

## 7. EXPERIMENTS RESULTS

We ran several experiments for each one of the previously discussed algorithms, trying to fine-tune their hyper-parameters to obtain solid results. Testing was organized in the following way. We generated two test sets (under the form of CSV files): one containing an actual labelled anomalous series of acquisitions that was provided to us by the people at ETC Riccione (Figure 5) and the other containing only artificial anomalies. The artificial anomalies were inserted into a series of legitimate acquisitions, by adding a random offset to a percentage of them (more precisely, 1% of the readings were altered by adding a random value between -4.0 and 4.0; no offsets between -1 and 1 were used because we considered them to be too subtle to be captured). Acquisitions to be altered were randomly sampled from the series.

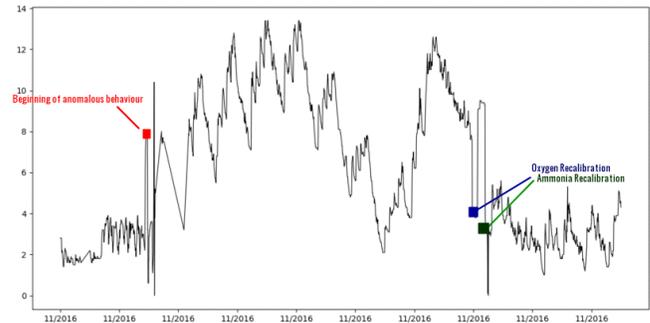

*Figure 5 Time series for ammonia acquisitions on trace 1 from the 9th of November to the 11th. No labels for single acquisitions were available (we only have a labelled window), however, we can pretty much consider any acquisition from the beginning of anomalous behaviour to the ammonia sensor calibration as an event.*

As a base of comparison, we designed a very naive anomaly detector. This baseline solution works by taking the integer component of the distance between each new measurement and the previous one. If the integer part of the distance is greater than 1, the new acquisition is registered as an anomaly. This solution is obviously prone to produce high false positive ratios. Figure 6 shows how the baseline classified each acquisition in the two test sets.

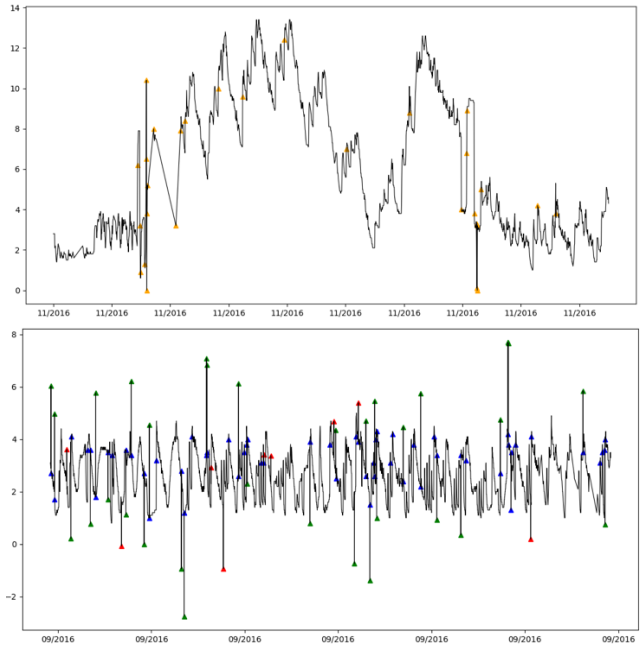

*Figure 6* Baseline results on the two test sets. On top, results on the documented anomaly. Orange triangles are readings classified as anomalous by the baseline. In the bottom picture, results on the synthetic test set. Green triangles are true positives, reds are false negative, blues are false positives.

As foreseeable, many false positives were generated by the baseline algorithm. On the synthetic test set, results were the following: 32 TP, 57 FP and 9 FN yielding a precision of 35.95% and recall of 78.04% and accuracy of 98.35%. Obviously, the accuracy measure is biased by the fact that only 1% of the readings are actually anomalies, therefore the very large number of true negatives pushes its value close to 100%. The value on which we'll evaluate the performance of algorithms on the synthetic dataset is instead the F1-measure, that for the baseline amounts to 49.23%.

Evaluating performance on the first test set is harder. Since we do not have per-acquisition labels, we can't compute precision, recall and F1-score there. Manually labelling all acquisitions inside the anomaly window as events wouldn't be optimal as well, as it would produce biased measures if we were to consider all those acquisitions as anomalies. Instead, we base our assessment on the frequency of anomalies inside the window. As we can see in Figure 6, the baseline solution did not produce any noticeable increment in the frequency of alarms during the anomaly period. This means that an event such as this one would have not been captured by the algorithm.

## 7.1 Low-High Pass Filter Results

Despite its simplicity, the low-high pass filter is highly customizable and therefore required several test runs to understand which configuration worked best. The algorithm, as already discussed in Section [Running Average Low-High Pass Filter], can be run either in online or offline mode, but always on univariate data. Since we were interested in detecting anomalies in the ammonia sensors, we obviously ran the algorithm on ammonia acquisitions, but also on the distance between ammonia and oxygen levels. We decided to try to use the difference in O2 and NH4 measurements as feature because we knew, from our previous analysis, that the two were correlated. From what the ETC experts told us in fact, under normal circumstances, high ammonia concentrations were accompanied by low oxygen levels and vice-versa. This meant that the distance between the two should pretty much remain the same during non-anomalous situations.

The two algorithm variants proved to behave similarly. In particular, we noticed that large window sizes for the computation of the running average, were necessary to identify anomalous trends such as that in the first test set (Figure 5). On the other side, to identify accurately sporadic outliers such as those that are present in test set two, a smaller window size was necessary. The configurations that gave the best results are the following:

- For the online version, a window size of 20 acquisitions and an alpha value of 0.2 worked best to identify sporadic anomalies in the second test set. We obtained a precision of 64.58%, a recall of 75.61% and a F1-score of 69.66%. However, to identify the anomalous trend of the first test set, a larger window of 1440 acquisitions was required, together with an alpha value of 0.5. Results are shown in Figure 7. As we can see, this time in the first test set a high frequency of anomalies was detected during the period of interest (highlighted in Figure 5), meaning that an incorrect behaviour would have been correctly detected.

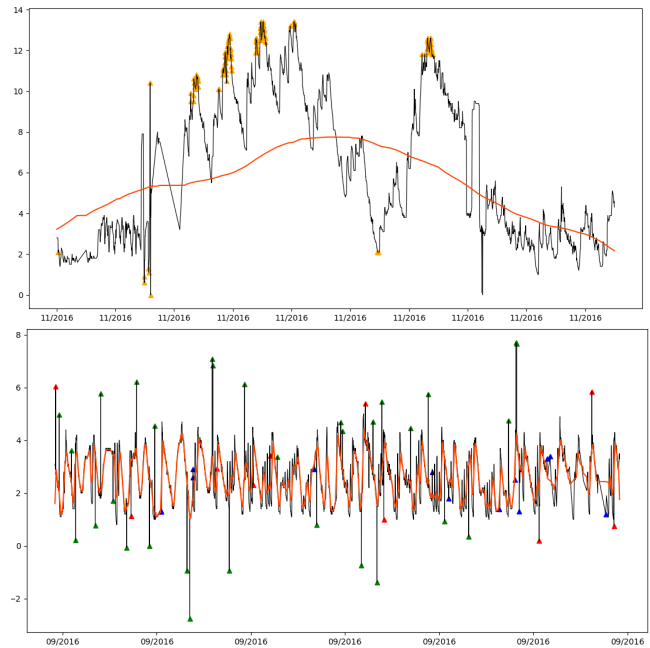

*Figure 7* Results with the online version of the high-low pass filter. The black line is the ammonia acquisition time series, the

*orange line is the running average. On the top image, the orange triangles are the anomalies identified by the algorithm. On the bottom image, green triangles are true positives (31), reds are false negatives (10), blues are false positives (17).*

- For the offline version, a window size of 5 was found to be better to identify sporadic anomalies in the second test set, together with an alpha value of 1, yielding a precision of 68.42% a recall of 95.12% and a F1-score of 79.59%. However, again, to detect an anomalous trend rather than a single anomalous acquisition, a larger window was necessary. In this case, we found that a window of size 600 (with alpha 1) produced reasonable results. For the offline versions, experiments are shown in Figure 8. Again, in the first test set, we can see how the frequency of anomalies spikes up during the period of interest.

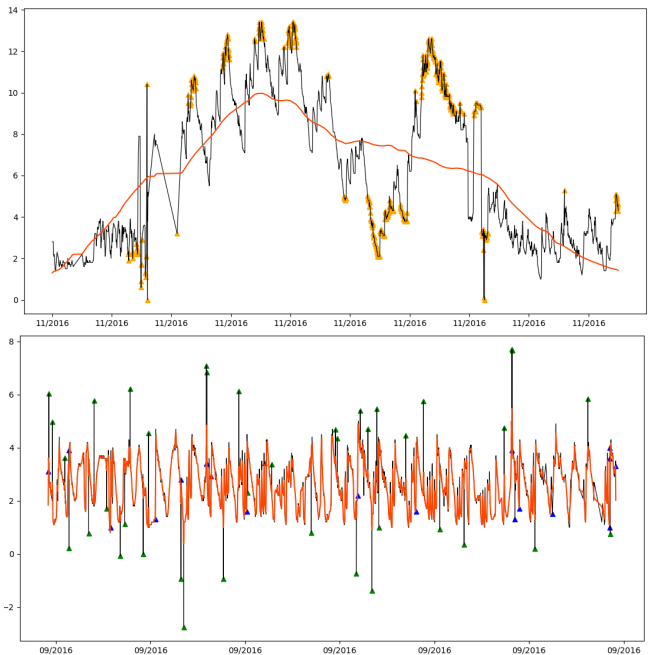

**Figure 8** Results with the offline version of the high-low pass filter. The black line is the ammonia acquisition time series, the orange line is the running average. On the top image, the orange triangles are the anomalies identified by the algorithm. On the bottom image, green triangles are true positives (39), reds are false negatives (2), blues are false positives (18).

Both versions of the algorithm proved to work better when ran on ammonia acquisitions, while our attempt of using the distance between O2 and NH4 levels always yielded inferior results. As expected, the offline version performed better, thanks to its access to the standard deviation of acquisitions.

The main problem of this algorithm is that it requires different settings depending on whether we are interested in capturing single anomalous acquisitions or anomalous trends in the time series. For this reason, we can conclude that the solution is not optimal. An idea to obtain a more solid algorithm would be to fuse two separate instances of the algorithm, one using settings to identify single outliers and one tuned to detect unusual trends and intelligently combine their output.

## 7.2 Univariate Gaussian Predictor Results

Tuning the univariate gaussian predictor did not take too long, since there was only a single hyper-parameter to set. As mentioned in Section [Univariate Gaussian Predictor], two different versions of the algorithm were designed: one that ran on single acquisitions and one that instead ran on windows of acquisitions. Both versions work identically, however, performance is easier to evaluate on the former. For this reason, we only discuss results obtained by said version of the algorithm.

Just like the low-high pass filter, the univariate gaussian predictor only works on mono-dimensional data that fits a Gaussian distribution. We already knew (as discussed in Section [Data Visualization]) that ammonia acquisitions met such a constraint. Additionally, we also found out that the distance between O2 and NH4 levels, also was approximately Gaussian, as shown in Figure 9. We therefore applied the algorithm on both the time series to see which one would give better results. Just like the low-high pass filter, the univariate gaussian predictor performed better when ran over ammonia acquisitions, rather than over O2 NH4 distances.

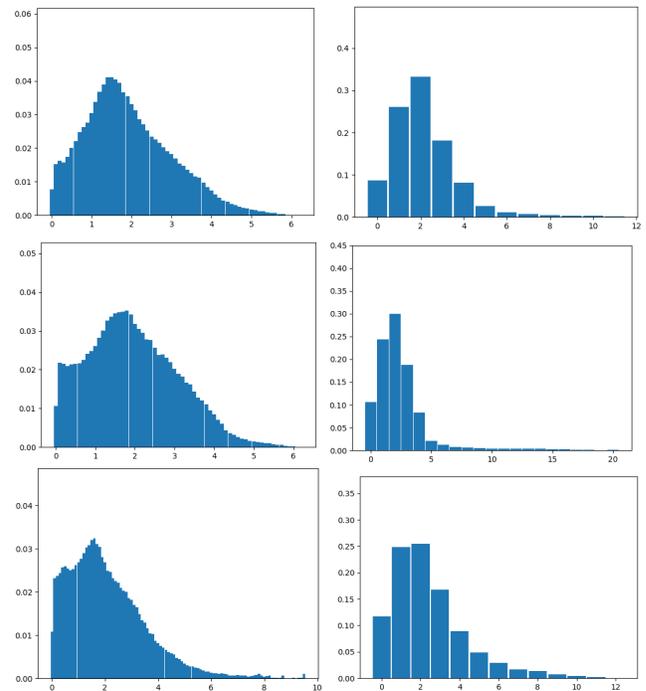

**Figure 9** Histograms of values' densities for O2 NH4 distances. The x axis contains the registered values, the y axis their densities. On the left we have densities of values rounded to 1 decimal digit. On the right densities for values rounded to integers. The first row is for first trace's sensors, the second row for second trace's sensors, the third row for third trace's sensors.

We trained our predictor over all the trace two's acquisitions (after they were cleaned as discussed in Section [Data Cleaning]) and then tested it over the usual two test sets. We tried out several values for the outlier threshold epsilon. The one that worked best was a value of 0.08, meaning that all acquisitions $x_i$ with probability $p(x_i) < 0.8\%$ would be registered as events. Figure 10 shows the performance of the predictor on both the test sets. As we can see, results are impressive for the first one, in which pretty much every acquisition inside the labelled window was marked as an anomaly. For the test set containing sporadic synthetic anomalies, the algorithm obtained a precision of 65.21%, a recall of 73.17% and a F1-score of 68.96%, which was less impressive.

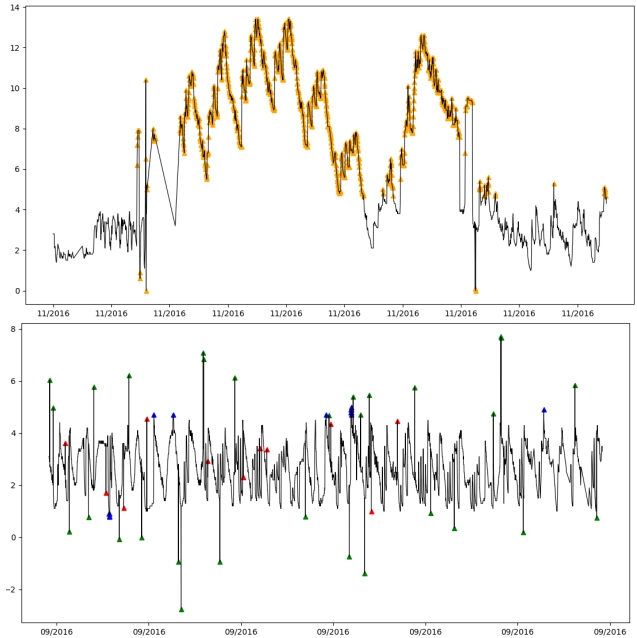

**Figure 10** Results for the univariate gaussian predictor over the two test sets. On top we have the dataset with the documented anomalous trend. Here, orange triangles mark acquisitions that were reported as anomalous. In the bottom picture we instead have performance on the second test set, with synthetic anomalies. Here, green triangles are true positives (30), reds are false negatives (11), blue are false positives (16).

The univariate gaussian predictor produced decent results overall, even though it had some issues in identifying smaller synthetic anomalies in the second test set. In order to be able to capture these as well, the algorithm requires a higher epsilon value, which in turn however, would generate many more false positives. We concluded that this behaviour can be explained by the fact that the probability of each measurement is computed in a context-less fashion, which may be too much of a simplifying assumption.

### 7.3 Seasonal-ESD Results

The Seasonal-ESD algorithm proved to be the best algorithm to identify sporadic outliers. In the second test set in fact, the algorithm obtained a precision of 96.96%, a recall of 78.04% and a F1-score of 86.48%. S-ESD however, did not perform well at all when used to identify anomalous trends. As we can see in Figure 11 in fact, almost no anomalies were detected in the window of interest. The only events registered by the algorithm took place at the beginning of the period, however, since they are fairly close in time, they may be enough to realize that something is wrong.

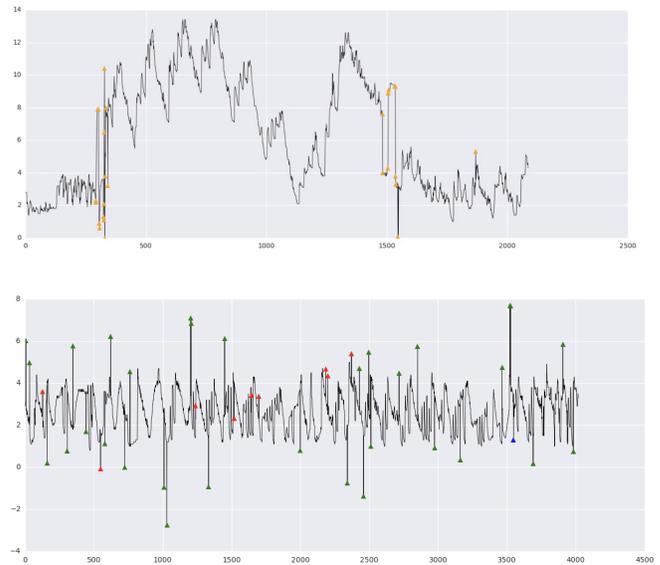

**Figure 11** Results for the Seasonal-ESD algorithm over the two test sets. On top we have the dataset with the documented anomalous trend. Here, orange triangles mark acquisitions that were reported as anomalous. In the bottom picture we instead have performance on the second test set, with synthetic anomalies. Here, green triangles are true positives (30), reds are false negatives (11), blue are false positives (16).

The main problem of this algorithm is obviously the fact that slow drifting of sensors acquisitions are not captured by it. This makes it inadequate to identify anomalous trends. As far as sporadic outliers are concerned however, this is the algorithm that yielded the best results.

### 7.4 LDCOF Results

LDCOF was the only algorithm, among the ones that we proposed, that could work with multiple features. This, together with the numerous hyper-parameters (, , the outlier score threshold and the number of clusters k), makes it very customizable and very difficult to tune. Testing every possible configurations would require an incredible amount of time. We therefore limited our explorations to those that made more sense.
Since increasing the dimensionality of samples too much hinders clustering performance and LDCOF is based on the k-means clustering algorithm, we decided to use only a small set of features consisting of: ammonia acquisitions, oxygen acquisitions and a temporal feature (either day of the week, month or season) to exploit the stationary property of the ammonia time series that was discussed in Section [Data Visualization] and the correlation between ammonia and oxygen levels. In Figure 12, we can see how feature selection impacts clustering.

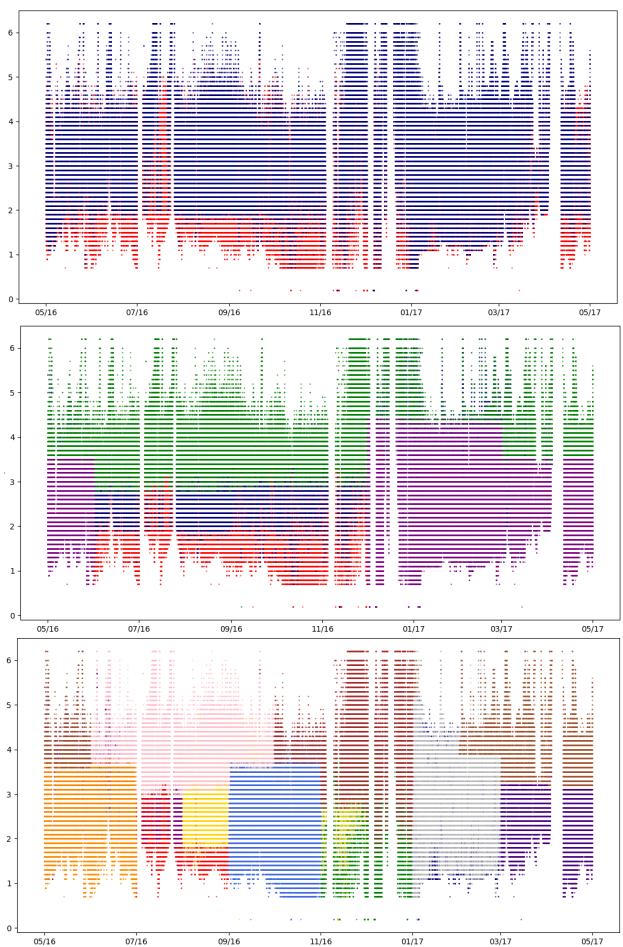

**Figure 12** Clusters obtained from ammonia acquisitions in trace two, when using: 2 clusters with ammonia and oxygen as features (top picture), 4 clusters with ammonia, oxygen and season as features (middle picture), 12 clusters with ammonia, oxygen and month as features (bottom picture).

After trying out several configurations for the algorithm, we obtained reasonable results on both datasets. LDCOF proved to be able to identify both anomalous trends and single outliers. The best results were obtained by running k-means with k = 12, = 0.75, = 0.25 and the outlier score threshold equal to the average distance of points in large clusters from their centroids, plus the standard deviation (samples that received a score above this threshold, were classified as outliers). As far as feature selection was concerned, we used ammonia acquisitions, oxygen acquisitions (since the two were correlated) and month of acquisitions (the reason why we selected a k = 12, was due to the fact that we were using months as features). Training (clustering) was performed on all the sensors' acquisitions of trace two. Results are shown in Figure 12. In the first test set, we can see how the frequency of registered anomalies increases dramatically during the period of interest. As far as the second test set is concerned, the algorithm was able to identify all major outliers, scoring a precision of 68.88%, a recall of 75.69% and a F1-measure of 72.09%. If we look at Figure 12 (bottom picture), we can see how only smaller anomalies were not captured.

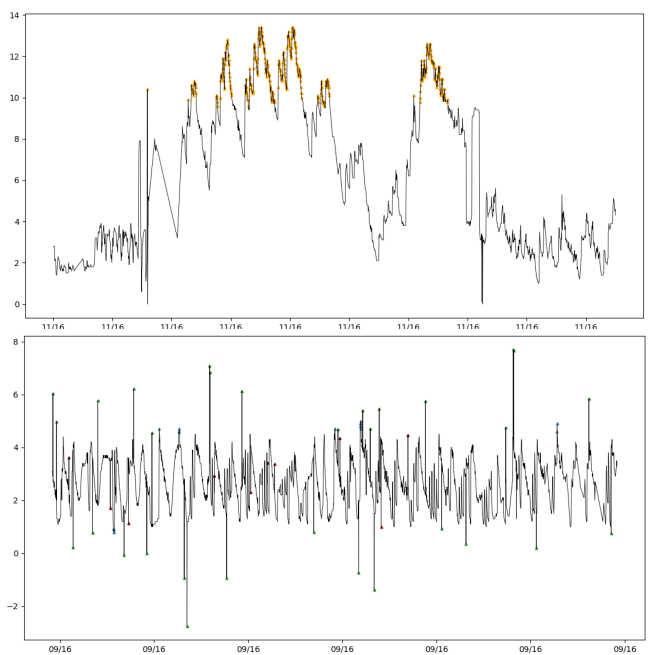

**Figure 13** Results of the LDCOF algorithms on both test sets. Above, we have the test set with the anomalous window of acquisitions that was provided to us by the experts at ETC. Orange triangles represent anomalies detected by the algorithm. Below, results over the dataset containing synthetic anomalies. Green triangles are true positives (31), reds are false negatives (10), blues are false positives (14).

The algorithm produced average results on both the test sets. LDCOF, together with the univariate gaussian predictor, were the only algorithms that performed well at detecting both anomalous acquisitions and anomalous trends. However, our opinion is that, between the two, the univariate gaussian predictor has the advantage due to its simplicity and fast training time.

## 8. CONCLUSIONS AND FURTHER WORK

Our experiments yielded very diverse results. In general, it is hard to identify which algorithm is better than the others overall, due to the fact that some perform better in identifying single outliers, while others are better in identification of anomalous trends. For this reason, an optimal solution could only be obtained by selecting a few of the proposed solutions to form a model based on an ensemble of experts. The experts' outputs would then be combined using either a majority vote approach, or a weight-based strategy, to decide which acquisition is to be classified as anomalous. Judging from our experiments, there are many reasonable combinations of experts. For instance, one may use two separate instances of the low-high pass filter, one tuned to detect single anomalies and one tuned to detect anomalous trends. Alternatively, one could use S-ESD to identify sporadic outliers, as this was the algorithm that performed better given such task and the univariate gaussian predictor to identify anomalous trends. As far as being able to detect sensor faults or damage from detected anomalies, we can only propose our solution, due to the fact that we did not have enough labelled anomalous trends (such as the one covering the period between the 9th and the 11th of November, that was used as test set) to test it. Our idea was to fix an interval t and then use the frequency of events that were usually registered in such interval during anomalous behaviours, to train a linear regressor that would be able to predict a frequency threshold T. Once we learned T, we would consider a sensor damaged whenever it was responsible for more than T detected anomalies in an interval t .